\documentclass[onecolumn]{mn2e}

\usepackage{times}
\input{psfig.sty}

\newif\ifAMStwofonts

\def\Mesz{M\'esz\'aros~}

\def\p{$e^\pm \;$}

\begin{document}

\title[Diversity of short GRBs] {On the diversity of short GRBs}
 
\author[Rosswog \& Ramirez-Ruiz]{Stephan Rosswog$^{1}$ and Enrico
        Ramirez-Ruiz$^{2}$ \\${\bf 1.}$ Department of Physics and
        Astronomy, University of Leicester, LE1 7RH, Leicester, UK.
        \\${\bf 2.}$ Institute of Astronomy, Madingley Road,
        Cambridge, CB3 0HA, UK.}

\date{}

\maketitle

\label{firstpage}

\begin{abstract}
Hydrodynamical simulations of the last inspiral stages and the final
coalescence of a double neutron star system are used to investigate
the power of the neutrino-driven wind, the energy and momentum of the
fireball produced via $\nu \bar{\nu}$-annihilation, and the 
intensity and character of their interaction. It is argued that 
the outflow that derives from the
debris will have enough pressure to collimate the relativistic
fireball that it surrounds. The low luminosity relativistic jet will
then appear brighter to an observer within the beam although most of
the energy of the event is in the unseen, less collimated and slower
wind. This model leads to a simple physical interpretation of the
isotropic luminosities implied for short GRBs at cosmological
distances. A wide variety of burst phenomenology could be attributable
to the dependence of the neutrino luminosity on the initial mass of
the double NS binary.
\end{abstract}


\section{Introduction}

Gamma-ray bursts (GRBs) come in two basic types with short ($\leq 2$
s) and long ($\geq 2$ s) duration (Kouveliotou et al. 1993). So far,
only the latter class has been relatively well-studied -- several {\it
long} bursts have been associated with afterglows that have been
observed from radio frequencies to X-ray energies (see \Mesz 2002 for
a recent review). By contrast recent X-ray, optical and radio searches
have not yet been successful for short GRBs (although see Lazzati,
Ramirez-Ruiz \& Ghisellini 2001). These bursts -- which make up about
one-third of all observed GRBs -- differ markedly from the long ones
not only in duration but also in having a larger fraction of
high-energy gamma rays in their energy distribution. The most widely
discussed possibility is that they result either from the merger of
two compact neutron stars or of a neutron star (NS) and a black hole
(BH; Lattimer \& Schramm 1976; Paczy\'{n}ski 1986, 1991; Eichler at
al. 1989; Narayan, Paczy\'{n}ski \& Piran 1992; Mochkovitch et
al. 1993; Klu\'{z}niak \& Lee 1998; Rosswog et al. 1999; Ruffert \&
Janka 1999; Lee \& Ramirez-Ruiz 2002; Rosswog \& Davies 2002).

The merger of two neutron stars produces a remnant with huge energy
reservoirs.  The difficult question to answer is how gravity and spin
can conspire to form the outflowing relativistic plasma 
that is required to explain the observations. The
conventional view is that the released energy is quickly and
continuously transformed into a radiation-dominated fluid with a high
entropy per baryon (Cavallo \& Rees 1978). This {\it fireball} is then
collimated into a pair of anti-parallel jets. There is as yet no
universally accepted explanation for jet collimation. The difficulty
of the problem is that there are interactions between the central
object (black hole or other), the inflow (i.e. disk), the outflow
(.i.e.wind) and the jet, and most of the dissension comes about
assessing condition and strength of these interrelationships
(e.g. Blandford 2002). Particular importance is attached here to the
interaction of the jet with the less collimated and slower outflow
associated with the disk. The most prominent feature of such
interaction is that the wind may be responsible for collimating the
jet (Levinson \& Eichler 2000; hereafter LE), and it is to this
problem that we have turned our attention.

We used three-dimensional high-resolution calculations of NS
coalescences to study the neutrino emission from the hot merger
remnant of a NS binary encounter. The issues we investigate here
include the properties of the neutrino-driven outflow (from the
debris), the energy and momentum injected into the region above the
remnant via $\nu \bar{\nu}$ annihilation, and the intensity and
character of their interaction. We estimate the dependence of the
neutrino luminosity on the initial mass of the double NS binary and
discuss the possible variety of afterglow behaviour that is expected
from fireballs that are collimated by a surrounding baryonic wind. Our
findings may be useful for designing search strategies for detecting
the afterglows of short GRBs.

\section{Neutrino emission in NS coalescence}

We have recently performed a series of three dimensional,
high-resolution simulations of the last inspiral stages and final
coalescence of a double NS using the smoothed particle
hydrodynamics method with up to $\approx 10^6$ particles (the reader
is referred to Rosswog \& Davies 2002 for further details on the
hydrodynamic evolution). Aware of its decisive role in the
thermo-dynamical evolution of the debris, we have used a {\it
realistic} equation of state for hot, dense nuclear matter, which is
based on the tables provided by (Shen et al. 1998a,b) and smoothly
extended to the low density regime with a gas consisting of neutrons,
alpha particles, photons and $e^{\pm}\;$ pairs.

Under the conditions encountered in the merger neutrinos are emitted
copiously and provide the most efficient cooling mechanism for the
dense, shock- and shear-heated debris. Moreover, the related weak
interactions determine the compositional evolution which is altered by
reactions such as electron and positron captures. The effect on the
cooling and the changes in the composition of the material are taken
into account via a detailed multi-flavour neutrino treatment which is
described in detail in Rosswog \& Liebend\"orfer (2003).  The results
presented here are based on the close analysis of late time segments,
typically a time $t_{\rm sim} \approx 15$ ms after the start of the
merger simulations.
%, where the neutrino luminosities have reached their
%maximum, stationary level. 
Results are described from three
representative runs: first, an initially corotating system
with twice 1.4 M$_{\odot}$ (c1.4), second, a system with twice 1.4
M$_{\odot}$ and no initial NS spins (i1.4), which we regard as the
generic case and finally, as an extreme case, a system of twice 2.0
M$_{\odot}$ and no initial spins (i2.0). In addition, in \S 4 a set 
of less resolved simulations is used to explore the dependence of the neutrino
luminosity on the system mass. 
 
The debris is heated initially via shocks (for example when two spiral
arms merge supersonically, see Fig. 13 in Rosswog \& Davies 2002) and
later via shear motion. The innermost part of the debris torus is
heated via shocks (to $T \approx 3$ MeV) that emerge when cool,
equatorially inflowing material collides with matter being shed from
the central object. The cool equatorial inflow ($T < 1$ MeV) allows
for the presence of heavy nuclei with a mass fraction of $\approx 10$
\% and mass numbers of $A \approx 80$ and $Z/A \approx 0.3$. Most
parts of the inner disk are essentially dissociated into neutrons and
protons. We find neutrino luminosities of $\approx 10^ {53}$ erg
s$^{-1}$ which are dominated by electron anti-neutrinos (Rosswog \&
Liebend\"orfer 2003). The mean energies are around 8, 15, and 22 MeV
for electron neutrinos, electron anti-neutrinos and the heavy lepton
neutrinos, respectively. Figure 1 shows the total neutrino energy per
time and volume in the orbital and meridional planes of the merged
remnant of our generic case (i1.4). The neutrino emission provides the
driving stresses that will lead to both a relativistic outflow
(through $\nu \bar{\nu}$-annihilation) and a less collimated (and
slower) outflow associated with the debris.

\subsection{$\nu\bar{\nu} \rightarrow e^{+}e^{-}$ above the remnant}

The neutrino annihilation process (which scales roughly with
$L_{\nu}^2$) can tap the thermal energy of the hot debris.  The
presence of a region of very low density along the rotation axis, away
from the equatorial plane of the debris must be present when the burst
takes place or otherwise prohibitive baryon loading will happen.  The
centrifugally evacuated funnel region above the remnant is an
attractive place for this deposition to occur because it is close to
the central energy source (the energy deposition rate scales roughly
with the inverse fourth power of the distance) but contains only a
small number of baryons (e.g. Davies et al. 1994, Ruffert et al. 1997,
Rosswog and Davies 2002).

The ratio of \p energy deposition to baryon rest mass energy,
$\eta=Q_{\nu\bar{\nu}}\tau_{\rm inj}/(\rho c^2)$, in the region above
the poles of the merged remnant is shown in Figure 1. $\rho$ denotes
the matter density and, for simplicity, an energy injection time,
$\tau_{\rm inj}= 1$ s has been assumed (Rosswog \& Ramirez-Ruiz
2002). The copiously emitted electron neutrinos and anti-neutrinos
dominate the annihilation process by $\approx$ 95 \%, their mean
average energies $\langle E \rangle$ being $\approx 8$ MeV for $\nu_e$
and $\approx 15$ MeV for $\bar{ \nu_{e}}$.

Due to the finite resolution of the simulations the densities along
the rotation axis are overestimated since they are mainly determined
by particles located in the inner parts of the disk. The above
restriction causes $\eta$ to be undervalued. Typical luminosities of
$\approx 10^{47}$ erg s$^{-1}$ (c1.4), $\approx 3 \times 10^{48}$ erg
s$^{-1}$ (i1.4), $\approx 10^{49}$ erg s$^{-1}$ (i2.0) are found in
the fireball region with $\eta > 1$. The largest attainable Lorentz
factors ($\approx 15$ although larger values are expected from higher
numerical resolution) are found along the binary rotation axis yet at
large angles away from it an increasing degree of entrainment leads to
a drastic decrease in $\eta$ (Fig. 1). The jet is likely to develop a
velocity profile so that different parts move with different Lorentz
factors. During the early stages of its evolution, this entrainment is
expected to give rise to a mixing instability which could provide
erratic changes in the relativistic jet velocity that would manifest
in internal shocks.

\begin{figure*}
%\centerline{\psfig{figure=comb.ps,angle=0,width=0.45\textwidth}
\centerline{\psfig{figure=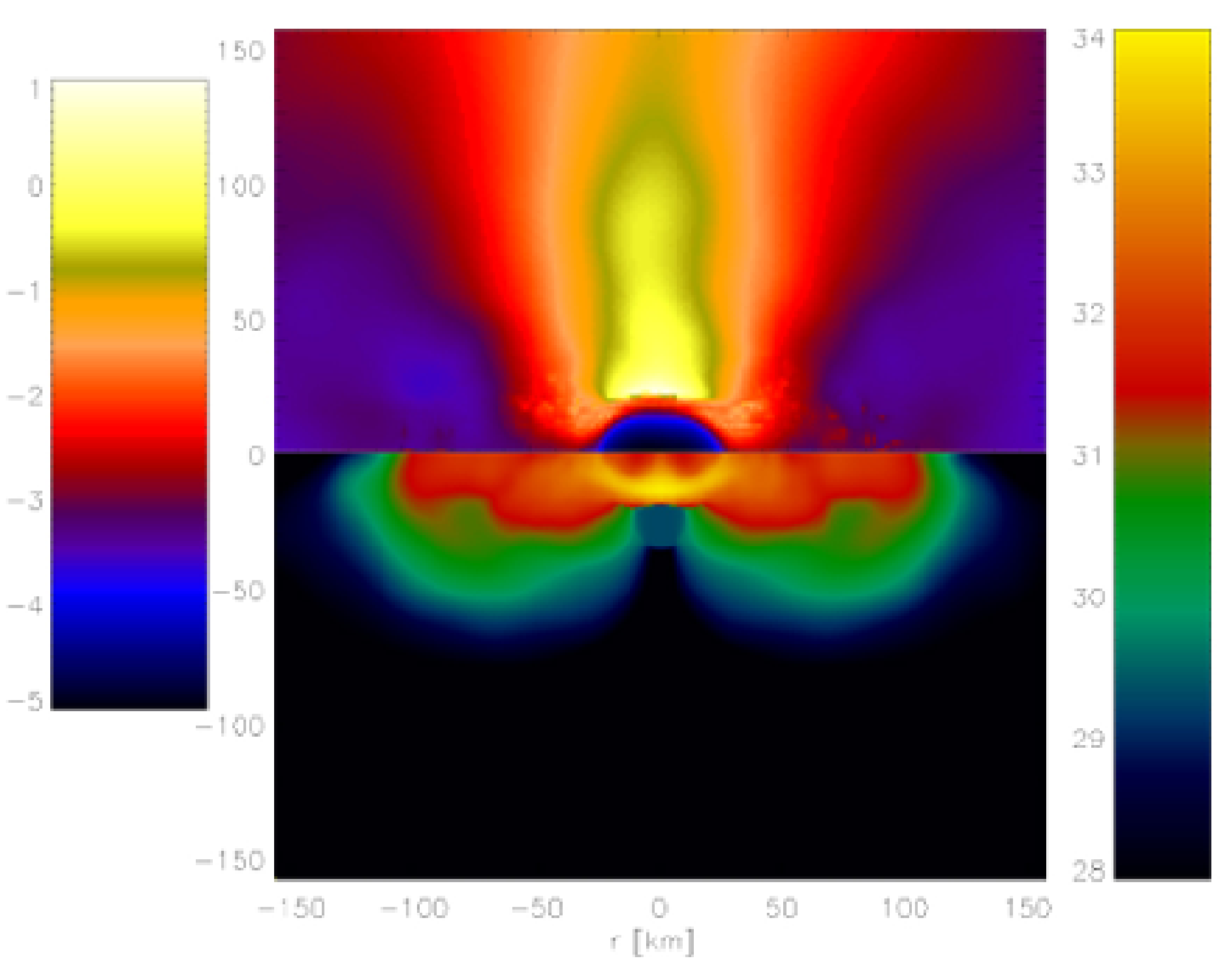,angle=0,width=0.49\textwidth}
\psfig{figure=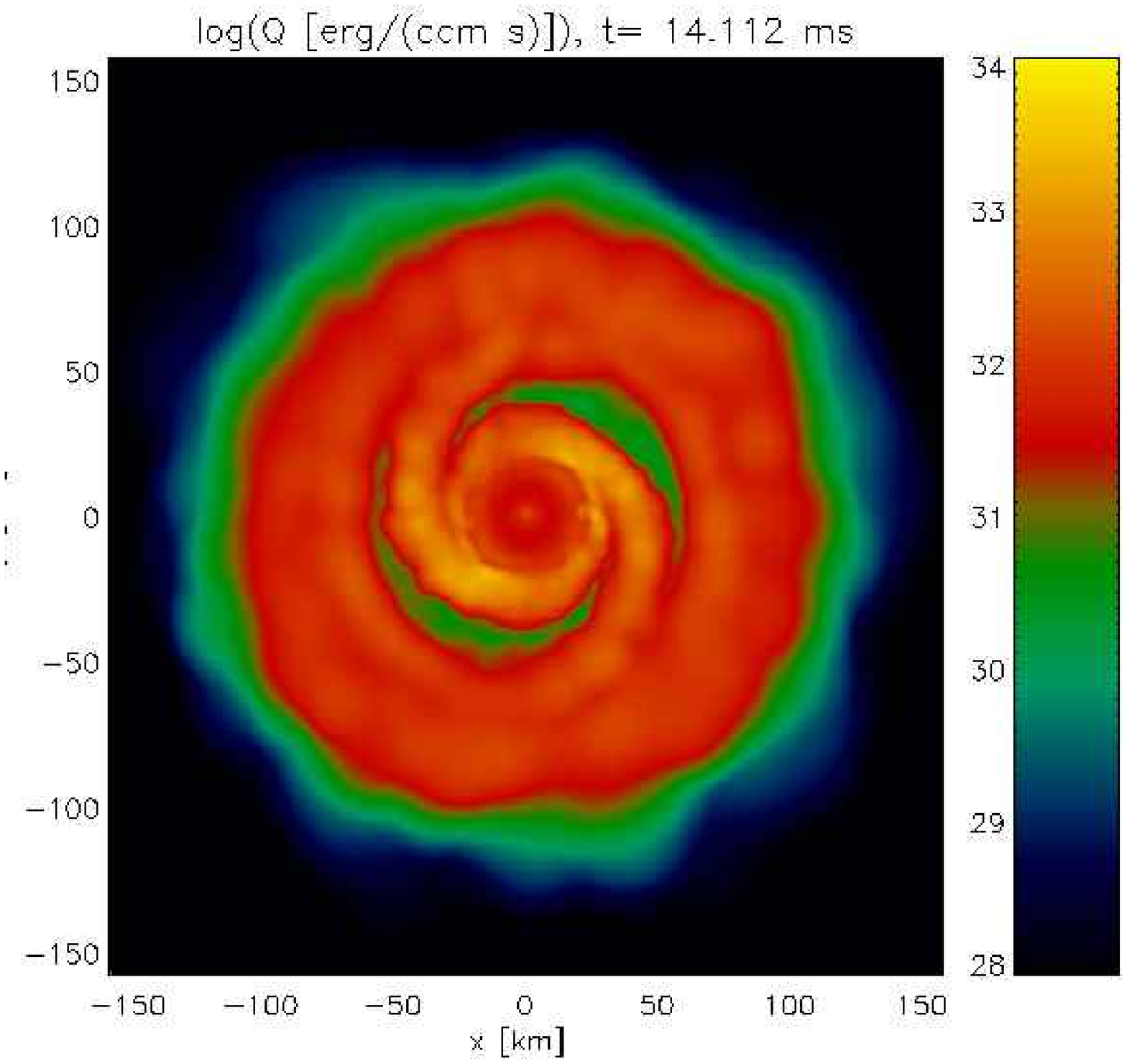,angle=0,width=0.42\textwidth}}
{\caption{Geometry of the neutrino  emission
above the merged remnant of model i1.4 (2x 1.4 M$_{\odot}$, no spin). 
The right panel shows the logarithm of the total neutrino energy 
emitted per time per volume (cgs units) from the orbital plane
while the left panel displays the meridional plane. 
The upper half-plane of the left panel shows the ratio of energy 
deposited during $\tau_{\rm inj}$ via $\nu\bar{\nu} \rightarrow e^{+}e^{-}$ 
to rest mass energy, $\eta$, which is a measure of the maximum attainable 
Lorentz factor. The colourbar on the left of this panel refers to 
$\log(\eta$). The lower half-plane of the left panel shows the total neutrino
emission in meridional plane. The colour coding is the same as in the right
panel.} 
\label{fig1}}
\end{figure*}

\subsection{Neutrino-driven wind}
The neutrinos that are emitted from the inner regions of the torus
will deposit part of their energy in the outer parts of the thick disk
that has formed around the central object. For example, a nucleon
located at $\approx 100$ km from the central object gains the
equivalent of its gravitational binding energy by capturing 3
neutrinos of $\approx 15$ MeV. Similar to the neutrino heating
mechanism in core-collapse supernovae, such neutrino deposition will
drive a strong mass outflow.  For high luminosities ($ L_{\nu}
>10^{52}$ erg s$^{-1}$), the mechanical power of the neutrino-driven
wind $L_w= 1/2 \dot{M} v^2_w$ (where $v_w$ is the asymptotic velocity)
is found to be largely independent of mass
(Thompson et al. 2001) with $L_w \propto L_\nu^{\alpha}$ and ${16
\over 5} < \alpha < {17 \over 5}$. Using Thompson et al. (2001) Table
1 we find $L_w= 2.0 \times 10^{49}$ erg s$^{-1}
L_{\nu,53}^{16/5}$. This yields a wind mechanical power of $\approx 2
\times 10^{49}$ erg s$^{-1}$, $\approx 2 \times 10^{50}$ erg s$^{-1}$,
and $\approx 2 \times 10^{51}$ erg s$^{-1}$ for runs c1.4, i1.4 and
i2.0.

While the above estimates agree with other estimates derived 
in the context of proto-neutron star formation (e.g. Duncan, 
Shapiro \& Wassermann 1986; Qian \& Woosley 1996)
they do not take into account the specific geometry of a merger 
remnant nor the extreme neutron richness of the
debris, we expect the corresponding correction factors to be of order
unity, which we consider accurate enough for the following discussion.

\section{The jet-wind connection}
We have now all the ingredients to consider the interaction of the jet
with the less collimated and slower outflow associated with the
debris. The most prominent feature of this interaction is
that the wind may be responsible for collimating the jet (LE). 
%It may also provide an effective invisible sheath that protects the rather
%fragile jet outflow from interaction with its environment. 
The collision of the two fluids will lead to the formation of a contact
discontinuity across which the total pressure is continuous. Besides,
two oblique shocks, one in each fluid, will form across where the
streamlines of the colliding (unshocked) fluids are deflected. The
structure of the shocked layers will depend on the parameters of the
two outflows and on the boundary conditions.

On the condition that the baryonic wind is highly supersonic not too
far out the streamlines of the colliding fluids, the momentum transfer
by the wind into the jet is dominated by ram pressure. The ram
pressure of the baryonic outflow just upstream the oblique shock at
the axial coordinate $z$ is then given by (LE)
\begin{equation}
(p_w + \rho_wc^2)U_w^2=\beta_w L_w/(4 \pi^2 c a_c r)\label{ram_p},
\end{equation}
where $p_w$, $\rho_wc^2$, and $U_w$ the corresponding wind pressure,
energy and four-velocity, $r$ is the distance between the torus and
the nearest point at the shock, $a_c(z)$ is the cross-sectional radius
of the contact discontinuity surface, and $\beta_w$ is the wind
terminal velocity (which is set to the escape velocity of the system
in the following estimates).

If the shocked wind layer is very thin due to rapid cooling of the gas
and using the requirement that the total pressure must be continuous
across the contact discontinuity, LE balance entropy and energy fluxes
to obtain: $\theta \approx \pi \beta_w^{-1} (L_j/L_w)$, where $\theta$
is the jet opening angle at large distances and the Lorentz factor at
the base of the jet has been assumed to be of order unity (this
approximation is justified in Rosswog \& Ramirez-Ruiz 2002).

In the above estimate $L_w$ is  the total wind power
emanating from a thin torus of radius $R$ centred around the central
object. This geometry is reminiscent of the neutrino emission
structure seen in Figure 1. Neutrinos emitted from the most luminous
ring-like region at about $R \approx 25$ km drive a baryonic outflow
whose total wind power surpass that ejected from rest of the debris by
at least an order of magnitude. Numerical integration of equation
(\ref{ram_p}), where the neutrino-driven wind is assumed to arise from
a sequence of concentric rings located at $z=0$, confirms that the
ring with the highest wind power determines the properties of the
cross-sectional radius of the jet at large distances ($z>R$). To first
order we therefore estimate $L_w$ to be the total wind power emanating
from this thin and most luminous region. In this way, at large enough
distances from the plane of the remnant, the semi-aperture angle of
the wind-confined jet is given by
\begin{equation}
\theta \approx 0.3\;\beta_{w,-1}^{-1}\;L_{w,50}^{-1}\;L_{j,48.5}, 
\end{equation}
where we adopt the convention $Q = 10^x\,Q_x$, using cgs units.  Once
the fireball Lorentz factor exceeds $\theta^{-1}$, it will remain
conical with the same opening angle regardless of the external
conditions. This occurs at a distance $\sim R \theta^{-1}$ from the
engine, which in turn implies that the baryonic outflow needs to
develop $t \sim 3 R_{1} \beta_{-1}^{-1} \theta_{-1}^{-1}$ ms before
the explosion that forms the GRB. This requirement is justified by the
fact that the neutrino luminosities reached their maximum, stationary
level $t > 15$ ms after the start of the simulations (see Fig. 2).

It is therefore likely that the neutrino-driven outflow that develops
in the merged debris will have enough pressure or inertia to provide
collimation. The beaming fraction is given by
$\varsigma=\Omega/2\pi\approx (\pi^2/2)(L_j/\beta_w L_w)^2$ and hence
the apparent luminosity is $L_\Omega \approx L_j \varsigma^{-1}= 2
\times 10^{49} \beta_{w,-1}^{2} L_{w,50}^{2}\;L_{j,48}^{-1}$ erg
s$^{-1}$. The above estimate gives $\theta \approx 0.1$ ($\approx
0.02$) and $L_\Omega \approx 10^{51}$ erg s$^{-1}$ ($L_\Omega \approx
3 \times 10^{52}$ erg s$^{-1}$) for runs i1.4 and i2.0 (assuming
$\alpha=16/5$), which clearly satisfies the apparent isotropized
energies of $\approx 10^{51}$ erg implied for short bursts at $z=1$
(Panaitescu et al. 2001; Lazzati et al. 2001).\\

The above reasoning does not take into account effects associated with
the formation of a rarefraction wave in the baryonic wind near the
interface separating the two fluids. This will depend on the
properties of the solution near the object and could result in a
steeper decline of the pressure supporting the jet, and the consequent
alteration of the profile at the contact discontinuity. There will
also surely be some entrainment of the electron-ion fireball plasma
and this should ultimately show up in the polarization observations
which can, in principle distinguish a pair plasma from a protonic
plasma. Entrainment will be promoted by linear instabilities that could
grow at the jet surface (Appl \& Camenzind 1993). In addition, there
can be an exchange of linear momentum with the surrounding outflow,
even if there is minimal exchange of mass. This can develop a velocity
profile in the jet so that different parts move with different Lorentz
factors.

\section{Diversity of afterglow behaviour}

The framework we have used to determine the geometry and energetics of
short GRBs is a simple one. Given the limitations of our assumptions
we draw three conclusions.  First, an inner relativistic jet
collimated by a surrounding baryonic wind can appear brighter to the
observer by a factor that is inversely proportional to its intrinsic
power (LE). Second, a modestly broad distribution of neutrino
luminosities can result in a wide variety of opening angles and
apparent luminosities. This can be understood as follows: the
luminosity of the $\nu\bar{\nu}$-annihilation jet increases as
$L_{\nu}^{2}$ while $L_w \propto L_{\nu}^{\alpha}$, so that $\theta
\propto L_{\nu}^{\alpha-2}$ and $L_{\Omega} \propto
L_{\nu}^{2\alpha-2}$ (${16 \over 5}<\alpha<{17 \over 5}$). Third, only
a small fraction of bursts are visible to a given observer with most
of the energy of the event being in the unseen outer outflow.

Figure 2 shows the results of a set of test calculations (with $\sim
6\times 10^4$ particles, no initial spin) performed in order to explore 
the effect that the masses of the neutron stars in the merging binary 
have on the total neutrino luminosity. 
\begin{figure*}
\centerline{\psfig{figure=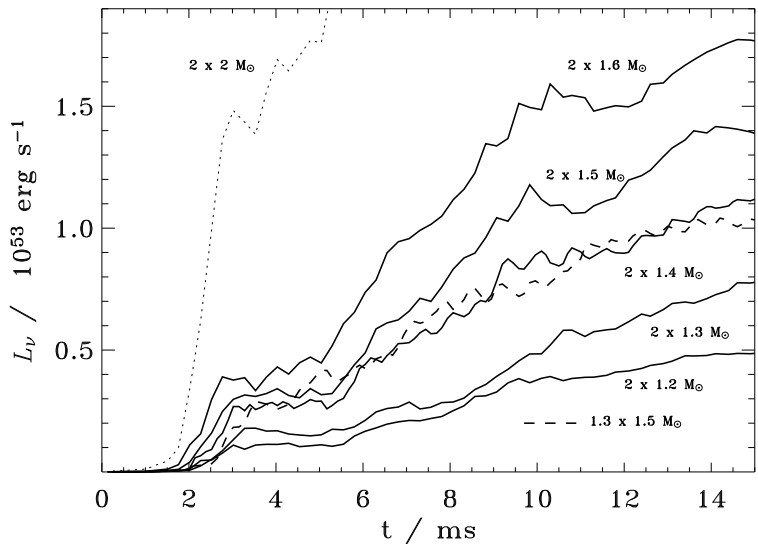,angle=0,width=0.7\textwidth}}
{\caption{Total neutrino luminosities as functions of time for various
NS/NS merger masses. The merger calculations were performed using a
SPH method with $\sim 6 \times 10^{4}$ particles.}
\label{fig2}}
\end{figure*}
From these models we derived an approximate relation between $L_\nu$
and the total mass of the compact binary, which we then apply to
estimate the possible variety of afterglow behaviour that could be
expected from NS-NS mergers.

We set up a $10^{6}$ grid of initial binary masses chosen according to
the theoretical distribution function of NS derived by Fryer \&
Kalogera (2001). The compact-remnant distribution is dominated by NSs
in the mass range 1.2-1.6$M_\odot$ and falls off drastically at higher
remnant masses (Fig. 3). Under the assumption that the total neutrino
luminosity (at late times) depends primarily on the total mass of the
compact binary (note that this is a reasonable assumption for mergers
of unequal masses provided that the binary mass ratio is not extreme;
see Fig. 2) and that $L_w \propto L_{\nu}^{\alpha}$ (Thompson et
al. 2001), we derive the distribution of opening angles and
luminosities expected from such encounters. Our results for two values
of $\alpha$ (16/5 and 17/5) are shown in Figure 3.

Since the jets are visible to only a fraction $\varsigma$
of the observers, the true GRB rate $R_{\tau}=\langle \varsigma
\rangle^{-1}R_{\rm obs}$, where $R_{\rm obs}$ is the observed rate and
$\langle \varsigma \rangle^{-1}$ is the harmonic mean of the beaming
fractions. We find $\langle \varsigma \rangle^{-1} \sim
100$. Following Schmidt (2001) $R_{\rm obs}(z=0)=0.5$ Gpc$^{-3}$
yr$^{-1}$. The true rate is then $R_{\tau}(z=0) \sim 50$ Gpc$^{-3}$
yr$^{-1}$, which should be compared with the estimated rate of NS
coalescence $R_{\rm NS}(z=0)\sim 80$ Gpc$^{-3}$ yr$^{-1}$ ($ \sim 2
\times 10^{-5}$ yr$^{-1}$ per galaxy; Phinney 1991). We note that the
simple collimation mechanism presented here provides an adequate
description of the observations.

\begin{figure*}
\centerline{\psfig{figure=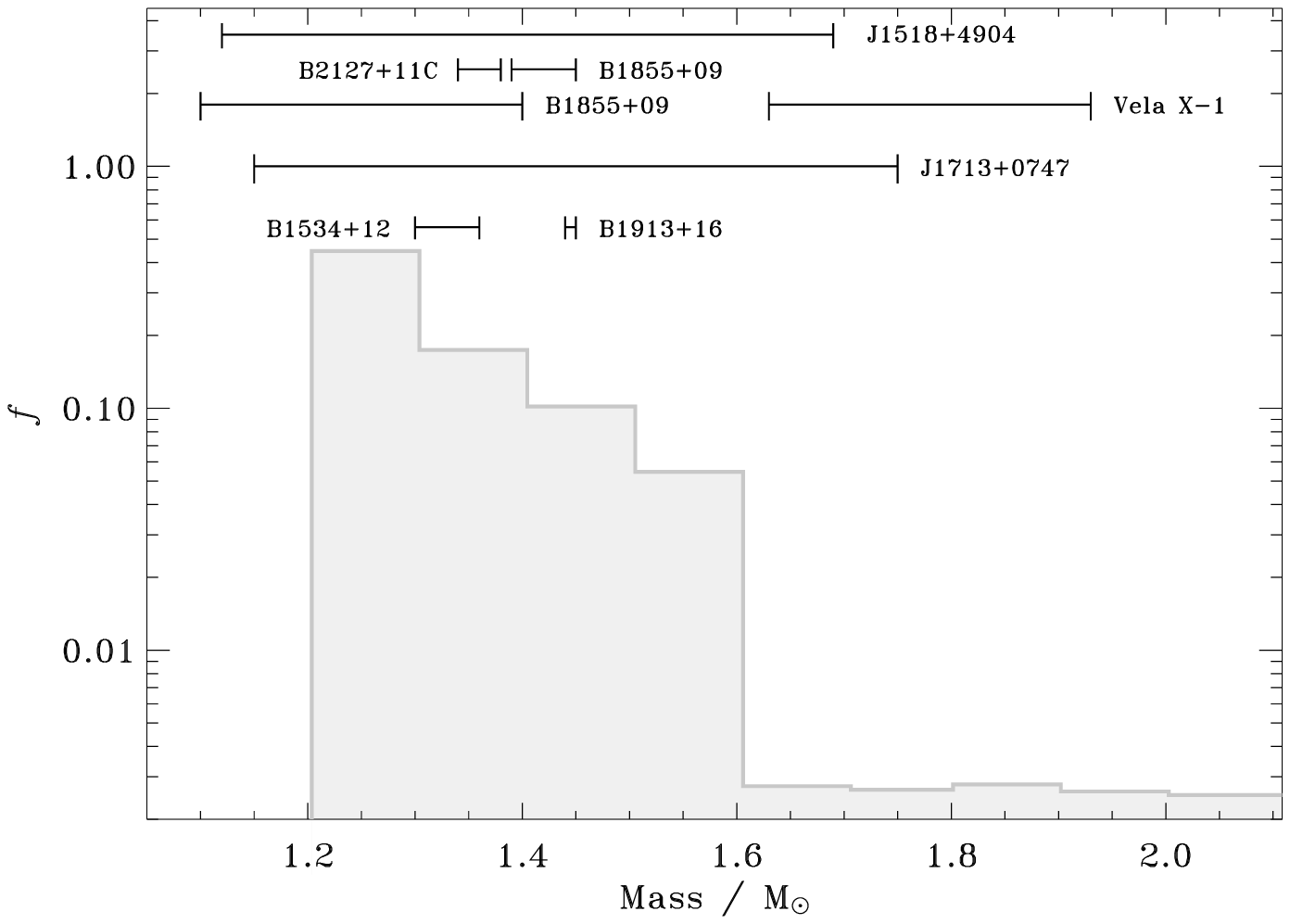,angle=0,width=0.54\textwidth}\psfig{figure=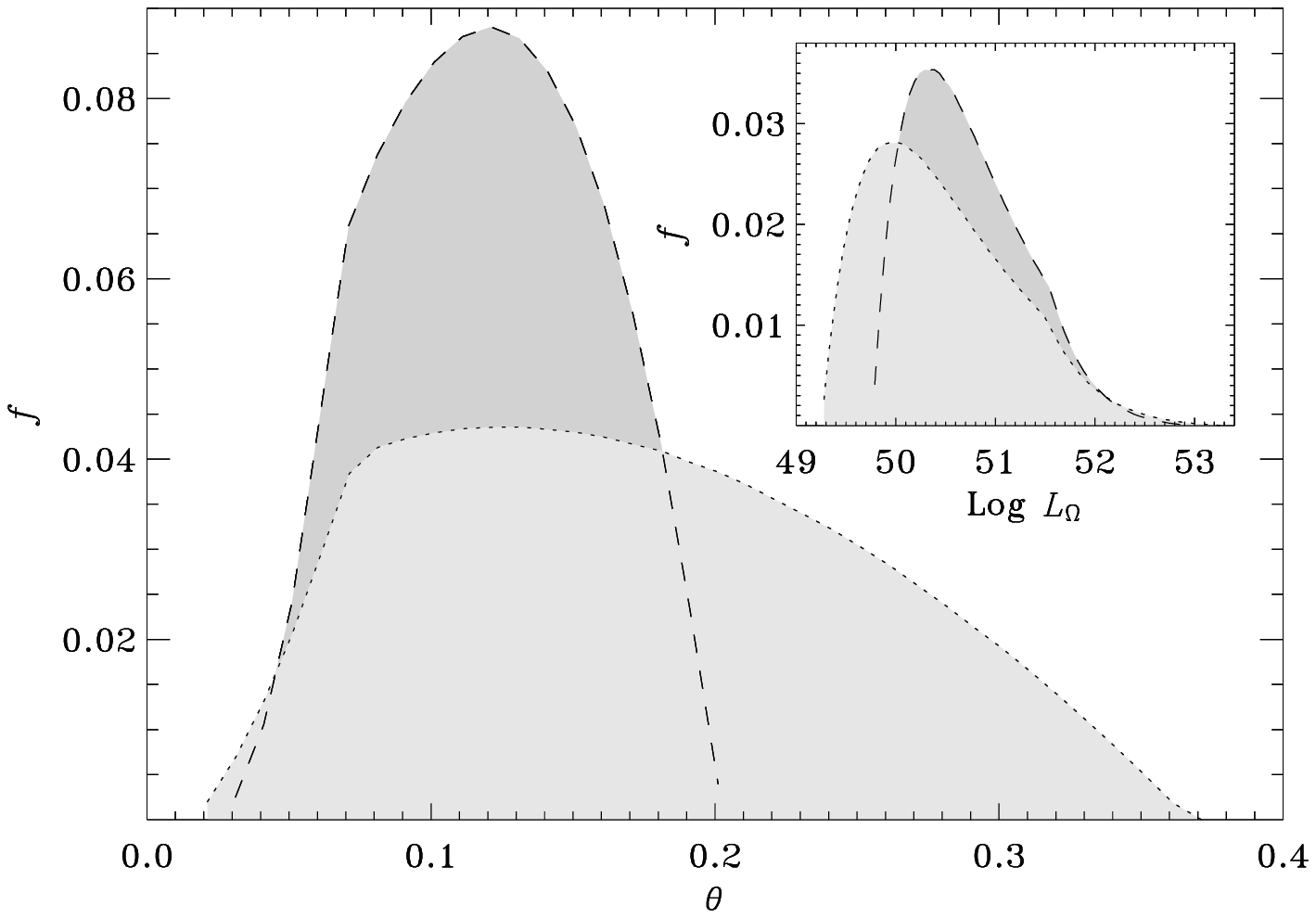,angle=0,width=0.54\textwidth}}
{\caption{Isotropic luminosities implied for short GRBs understood in
terms of collimation of the fireball by an outer baryonic outflow.The
left panel shows the distribution of neutron-star remnant masses
derived by Fryer \& Kalogera (2001). The NS masses are measured by
Thorsett \& Chakrabarty (1999) and Barziv et al.  (2001). The right
panel shows the distribution of expected opening angles and
luminosities for NS binaries. The dashed line assumes $L_w \propto
L_{\nu}^{16/5}$ while the dotted line uses $L_w \propto
L_{\nu}^{17/5}$.}
\label{fig3}}
\end{figure*}

\section{Summary}

In this letter we consider the viability of NS-NS binary coalescences 
as central engines of short GRBs. The main form of energy release we 
regard is neutrino emission from the disk\footnote{More energy 
could be pumped into the
$e^{\pm}\gamma$ fireball when the rapidly rotating BH is formed or if
magnetics field are able to tap the rotational energy of the BH with
higher efficiency than $\nu\overline{\nu}$ does (Popham, Woosley \&
Fryer 1999; Rosswog, Ramirez-Ruiz \& Davies 2003).}.

We suggest that the isotropic luminosities implied for short GRBs at
cosmological distances can be understood in terms of collimation of
the $\nu\overline{\nu}$-annihilation fireball by the neutrino-driven
wind. We argue that the existence of such a baryonic wind is a natural 
consequence of the huge gravitational binding energy released in the form of
neutrinos.

Moreover, we show that the wind derived from the disk could in
principle have enough pressure or inertia to provide collimation to
the fireball that derives from the region around the BH. In other
words, bursts produced by such a mechanism are not isotropic but
instead beamed within a solid angle, typically $\sim 0.1$
sterad. Within the framework of this simple model we have deduced the
distribution of opening angles and luminosities expected from
$\nu\bar{\nu}$-annihilation in NS-NS mergers and uncovered that within
the uncertainties of the calculations, the coalescence scenario is
capable of providing the required isotropized energies and a
compatible number of progenitors.

Given the few hundred kilometres per second acquired by the NS at
birth and assuming a typical coalescence time of about 100 Myr, these
bursts should occur predominantly in the low-density halo of the
galaxy. The external shock can therefore occur at much larger radii
and over a much longer timescale than in usual afterglows, and the
X-ray intensity is below the threshold for triggering. Alternatively,
the burst could go off inside a pulsar cavity inflated by one of the
neutron stars in the precursor binary. Such cavities can be as large
as fractions of a parsec or more, giving rise to a deceleration shock
months after the GRB with a consequently much lower brightness that
could avoid triggering and detection. In conclusion, the absence of
detected afterglows in short bursts is not surprising, and, as argued
above, a wide diversity of behaviours may be the rule, rather than the
exception.

\section*{Acknowledgements}
It is a pleasure to thank M. J. Rees \& A. MacFadyen for discussions
and the Leicester supercomputer team S.  Poulton, C. Rudge and R. West
for their excellent support.  The computations reported here were
performed using both the UK Astrophysical Fluids Facility (UKAFF) and
the University of Leicester Mathematical Modelling Centre's
supercomputer. This work was supported by PPARC, CONACyT, SEP and the
ORS foundation.

\end{document}